\documentclass{article}
\usepackage{amssymb}
\usepackage{amsmath}

\begin{document}

\title{Physical propositions and quantum languages}
\author{Claudio Garola\thanks{Dipartimento di Fisica dell'Universit\`{a} e Sezione INFN, 73100 Lecce,
Italy; e-mail: garola@le.infn.it.}}
\maketitle

\begin{abstract}
The word \textit{proposition} is used in physics with different meanings, which must be distinguished to avoid interpretational problems. We construct two languages $\mathcal{L}^{\ast}(x)$ and $\mathcal{L}(x)$ with classical set-theoretical semantics which allow us to illustrate those meanings and to show that the non-Boolean lattice of propositions of quantum logic (QL) can be obtained by selecting a subset of \textit{p-testable} propositions within the Boolean lattice of all propositions associated with sentences of $\mathcal{L}(x)$. Yet, the aforesaid semantics is incompatible with the standard interpretation of quantum mechanics (QM) because of known no-go theorems. But if one accepts our criticism of these theorems and the ensuing SR (semantic realism) interpretation of QM, the incompatibility disappears, and the classical and quantum notions of truth can coexist, since they refer to different metalinguistic concepts (\textit{truth} and \textit{verifiability according to QM}, respectively). Moreover one can construct a quantum language $\mathcal{L}_{TQ}(x)$ whose Lindenbaum-Tarski algebra is isomorphic to QL, the sentences of which state (testable) properties of individual samples of physical systems, while standard QL does not bear this interpretation.
\end{abstract}

\section{Introduction}
The word \textit{proposition} has been used in physics with some different meanings. Jauch (1968) intended it simply as a synonim of \textit{yes-no experiment}, Piron (1976) denoted by it an equivalence class of \textit{questions}, etc., following a tradition started by Birkhoff and von Neumann (1936) with their \textit{experimental propositions}. On the other hand, the same term is also used in order to denote the (closed) set of states associated with an experimental proposition, often called \textit{physical proposition} (see, e.g., Dalla Chiara \textit{et al}., 2004, Introduction to Part I). The latter use is commonly preferred by those logicians concerned with quantum logic (QL) who identify states with \textit{possible worlds }(\textit{ibid.}, Ch. 8). For, an experimental proposition can be considered as a sentence of a physical language, and the set of states associated with it as its proposition in a standard logical sense. However, the term \textit{proposition} is also used to denote an element of the Lindenbaum-Tarski algebra of the aforesaid physical language (see, e.g., R\'{e}dei, 1998, Ch. 5; the links between the two meanings are rather obvious).

Let us adopt from now on the standard logical meaning of the term proposition, accepting to identify physical states with possible worlds (which may be questioned from several viewpoints; we, however, do not want to discuss this topic in the present paper). Then, a serious problem occurs when dealing with quantum mechanics (QM), hence with QL. Indeed, every
Birkhoff and von Neumann's experimental proposition can be experimentally
confirmed or refuted (see also Jammer, 1974, Ch. 8), so that it can be
interpreted as a sentence $\alpha $ of an observative language, stating a
physical property that can be tested on one or more individual samples of a
given physical system (\textit{physical objects}). In classical mechanics
(CM) a truth value is defined for every (atomic or molecular) sentence $%
\alpha $, and the physical proposition p$_{\alpha }$ of $\alpha $ (meant as
a set of states in which $\alpha $ is true) is introduced basing on this
definition. On the contrary, it can occur in QM that no truth value can be
defined for a sentence $\alpha $ because of \textit{nonobjectivity of
properties} (equivalently, the distinction between \textit{actual }and 
\textit{potential} properties), which is a well known and debated feature of
this theory (see, e.g., Busch \textit{et al.}, 1991, Ch. II; Mermin, 1993).
Indeed, nonobjectivity prohibits one to associate a physical property $E$
with a set of physical objects possessing $E$, which is a basic step if one
wants to construct a classical set-theoretical semantics. Hence, a physical
proposition is directly associated, in QM, with $\alpha $, whose truth value
is defined via the proposition itself. This gives rise to a number of
difficulties, since the notion of truth introduced in this way has several
odd features. For instance, if a sentence is not true in a possible world
(state), one cannot assert that it is false in that state, and the join of
two sentences may be true even if none of the sentences is true. More
important, this notion of truth clashes with the fact that every
(elementary) experimental proposition can be checked on a physical object,
yielding one of two values (0 or 1) that can be intuitively interpreted as 
\textit{true} and \textit{false}. Thus, the identification of sentences with
their propositions may produce serious troubles (the ``metaphysical
disaster'' pointed out, though in a somewhat different way, by Foulis and
Randall, 1983). According to Dalla Chiara \textit{et al.} (2004, Ch. 1) this
problem stimulated the investigation about more and more general quantum
structures. In our opinion, however, the attempt at solving it in this way
is questionable. Indeed, the problem is originated by some specific features
of the standard interpretation of the mathematical formalism of QM (to be
precise, the aforesaid nonobjectivity of properties) and not by the
formalism itself, so that it cannot be solved by simply generalizing the
mathematical apparatus without removing those peculiarities of the
interpretation that create it (see also Busch and Shimony, 1996).

According to a widespread belief, the impossibility of solving the above
problem by firstly endowing the language of QM with a classical
set-theoretical semantics and then introducing the set of propositions is
witnessed by the fact that this set has a structure of orthomodular
nondistributive lattice, while a classical semantics would lead to a Boolean
lattice of propositions.

We aim to show in this paper that the above belief is ill-founded. To be
precise, we want to show that one can construct a simple language $\mathcal{L%
}(x)$ endowed with a classical set-theoretical semantics, associate it with
a poset of \textit{physical propositions} (that generally is not a lattice),
and then introduce a definition of \textit{testability} on $\mathcal{L}(x)$
which selects a subposet of \textit{testable }(actually, \textit{p-testable}%
, see Sec. 3) \textit{physical propositions}. Our procedure is very
intuitive, and applies to every theory, as CM and QM, in which physical
objects and properties can be defined. Under reasonable physical assumptions
the poset of all testable physical propositions turns out to be a Boolean
lattice in CM, while it is an orthomodular nondistributive lattice in QM
that can be identified with a (standard, sharp) QL. It follows, in
particular, that nondistributivity cannot be considered an evidence that a
classical notion of truth cannot be introduced in QM.

Our result does not prove, of course, that providing a classical semantics
for the observative language of QM is actually possible. Indeed,
nonobjectivity of properties would still forbid it. However, should one
accept the criticism to nonobjectivity provided by ourselves in some
previous paper, and the \textit{Semantic Realism} (SR) interpretation of QM
following from it (see Garola and Solombrino, 1996a, 1996b; Garola 1999,
2000, 2002, 2005; Garola and Pykacz, 2004),\footnote{%
We remind that our criticism is based on an epistemological perspective
according to which the \textit{theoretical laws} of any physical theory are
considered as mathematical schemes from which\textit{\ empirical laws} can
be deduced. The latter laws are assumed to be valid in all those physical
situations in which they can be experimentally checked, while no assumption
of validity can be done in physical situations in which some general
principle prohibits one to check them (this position is consistent, in
particular, with the operational and antimetaphysical attitude of standard
QM). In CM our perspective does not introduce any substantial change, since
there is no physical situation in which an empirical law cannot, in
principle, be tested. On the contrary, if boundary, or initial, conditions
are given in QM in which properties that are not compatible are attributed
to the physical system (more precisely, to a sample of it), a physical
situation is hypothesized that cannnot be empirically accessible, hence no
assumption of validity can be done for the empirical laws deduced from the
general formalism of QM in this situation. Strangely enough, this new
perspective is sufficient to invalidate the proof of some important no-go
theorems, as Bell's (Bell, 1964) and Bell-Kochen-Specker's (Bell, 1966;
Kochen and Specker, 1967). Nonobjectivity of properties then appears in this
context as an interpretative choice, not a logical consequence of the
theory, and alternative interpretations become possible. Among these, our SR
interpretation restores objectivity of properties without requiring any
change in the mathematical apparatus and in the minimal (statistical)
interpretation of QM.} the language $\mathcal{L}(x)$ introduced in this
paper appears as a sublanguage of the broader observative language of QM,
and the classical set-theoretical semantics defined on it can be seen as a
restriction of the broader classical set-theoretical semantics that can be
defined on the observative language. If this viewpoint is accepted, the
distinction between physical propositions and testable physical propositions
can be considered something more than an abstract scheme for showing how
non-Boolean algebras can be recovered within a Boolean framework. Indeed,
physical propositions are then associated in a standard way with
(universally) quantified sentences of $\mathcal{L}(x)$ that have classical
truth values, which avoids the ``metaphysical disaster'' mentioned above,
and testable physical propositions are physical propositions associated with
quantified sentences for which \textit{truth criteria} are given that allow
one to determine empirically their truth values.\footnote{%
From a logical viewpoint our treatment exhibits the deep reasons of the
``disaster''. Indeed, \textit{experimental} propositions are interpreted as
open sentences of a first order predicate language, while \textit{physical}
propositions are associated with quantified sentences of the same language.}

The lattice operations on the lattice of all testable physical propositions,
however, only partially correspond to logical operations of $\mathcal{L}(x)$
in QM. We show that $\mathcal{L}(x)$ can be enriched by introducing new 
\textit{quantum connectives}, so that a language $\mathcal{L}_{TQ}(x)$ of
testable sentences can be extracted from $\mathcal{L}(x)$ whose
Lindenbaum-Tarski algebra is isomorphic to the orthomodular lattice of all
testable physical propositions of $\mathcal{L}(x)$. Thus, we introduce a
clear distinction between classical and quantum connectives, and show that a
verificationist notion of \textit{quantum truth} can be defined on $\mathcal{%
L}_{TQ}(x)$ which coexists with the classical definition of truth, rather
than being alternative to it. This is a noticeable achievement, which avoids
postulating that different incompatible notions of truth are implicitly
introduced by our physical reasonings.

Some of the results resumed above have already been expounded in some
previous papers (Garola and Sozzo, 2004, 2006), though in a somewhat
different form. Here we generalize our previous treatments by considering
effects in place of properties, which leads us to preliminarily construct a
broader language $\mathcal{L}^{\ast }(x)$ in which $\mathcal{L}(x)$ is
embedded. An interesting consequence of this broader perspective is a
weakening of the notion of testability, which illustrates from our present
viewpoint a possible advantage of unsharp QM with respect to standard QM. We
also provide a simple new way for defining physical propositions by
introducing universal quantifiers on the sentences of the language $\mathcal{%
L}^{\ast }(x)$, which also helps in better understanding the notion of
quantum truth and its difference from classical truth. For the sake of
brevity, however, our presentation is very schematic and essential.

It remains to observe that a more general treatment of the topics discussed
in this paper could be done by adopting the formalization of an observative
sublanguage of QM introduced by ourselves many years ago (Garola, 1991). In
this case, two classes of predicates would occur, one denoting effects
(hence properties), one denoting states, so that states would not be
identified with possible worlds and physical propositions would be
distinguished from propositions in a standard logical sense. This treatment
would be more general and formally complete, at the expense, however, of
simplicity and understandability, so that we do not undertake this task here.

\section{The language of effects $\mathcal{L}^{\ast }(x)$}

We call $\mathcal{L}^{\ast }(x)$ the formal language constructed by means of
the following symbols and rules.

\smallskip

\textit{Alphabet.}

An individual variable $x$.

Monadic predicates $E$, $F$, $...$.

Logical connectives $\lnot $, $\wedge $, $\vee $.

Auxiliary signs $($, $)$.

\smallskip

\smallskip \textit{Syntaxis.}

Standard classical formation rules for well formed formulas (briefly, 
\textit{wffs}).

\smallskip

We introduce a \textit{set-theoretical semantics} on $\mathcal{L}^{\ast }(x) 
$ by means of the following metalinguistic symbols, sets and rules.

$\mathcal{E}^{\ast }$: the set of all predicates.

$\Phi ^{\ast }(x)$: the set of all wffs of $\mathcal{L}^{\ast }(x)$.

$\mathcal{E}^{\ast }(x)$: the set $\{E(x)\mid E\in \mathcal{E}^{\ast }\}$ of
all \textit{elementary} wffs of $\mathcal{L}^{\ast }(x)$.

A set $\mathcal{S}$ of \textit{states}.

For every $S\in \mathcal{S}$, a universe $\mathcal{U}_{S}$ of \textit{%
physical objects}.

A set $\mathcal{R}$ of mappings (\textit{interpretations}) such that, for
every $\rho \in \mathcal{R}$, $\rho :(x,S)\in \{x\}\times \mathcal{%
S\longrightarrow \rho }_{S}(x)\in \mathcal{U}_{S}$.

For every \textit{S}$\in \mathcal{S}$ and $E\in \mathcal{E}^{\ast }$, an 
\textit{extension} $ext_{S}(E)\subseteq $ $\mathcal{U}_{S}$.

For every $\rho \in \mathcal{R}$ and $S\in \mathcal{S}$, a \textit{classical
assignment function }$\sigma _{S}^{\rho }:\Phi ^{\ast }(x)\longrightarrow
\{t,f\}$ (where $t$ stands for \textit{true} and $f$ for \textit{false}),
defined according to standard (recursive) truth rules in Tarskian semantics
(to be precise, for every elementary wff $E(x)\in \mathcal{E}^{\ast }(x)$, $%
\sigma _{S}^{\rho }(E(x))=t$ iff $\rho _{S}(x)\in ext_{S}(E)$, for every
pair $\alpha (x)$, $\beta (x)$ of wffs of $\Phi ^{\ast }(x)$, $\sigma
_{S}^{\rho }(\alpha (x)\wedge \beta (x))=t$ iff $\sigma _{S}^{\rho }(\alpha
(x))=t=$ $\sigma _{S}^{\rho }(\beta (x))$, etc.).

\smallskip

The \textit{intended physical interpretation} of $\mathcal{L}^{\ast }(x)$
can then be summarized as follows.

Reference to a physical system $\Sigma $ is understood.

A predicate of $\mathcal{L}^{\ast }(x)$ denotes an \textit{effect}, which is
operationally interpreted as an equivalence class of (dichotomic) \textit{%
registering devices}, each of which, when activated by an individual sample
of $\Sigma $, performs a \textit{registration} that may yield value 0 or 1
(see, e.g., Ludwig, 1983, Ch. II; Garola and Solombrino, 1996). We assume in
the following that every registering device belongs to an effect.

A state is operationally interpreted as an equivalence class of \textit{%
preparing devices}, each of which, when activated, performs a \textit{%
preparation} of an individual sample of $\Sigma $ (\textit{ibid.}).

A physical object is operationally interpreted as an individual sample of $%
\Sigma $, which can be identified with a preparation (\textit{ibid.}).

The equation $\sigma _{S}^{\rho }(E(x))=t$ (or $f$) is interpreted as
meaning that, if a registering device belonging to $E$ is activated by the
physical object $\mathcal{\rho }_{S}(x)$, the result of the registration is
1 (0). The interpretation of $\sigma _{S}^{\rho }(\alpha (x))=t$ (or f),
with $\alpha (x)\in \Phi ^{\ast }(x)$, follows in an obvious way, bearing in
mind the above truth rules for the connectives $\lnot ,\wedge ,\vee $.

\smallskip

Let us now introduce some further definitions and notions.

(i) We define a\textit{\ logical preorder} $<$ and a \textit{%
logical equivalence} $\equiv $ on $\Phi ^{\ast }(x)$ in a standard way, as
follows.

Let $\alpha (x)$, $\beta (x)\in \Phi ^{\ast }(x)$. Then,

$\alpha (x)<\beta (x)$\quad \textit{iff}\quad for every $\rho \in \mathcal{R}
$ and $S\in \mathcal{S}$, $\sigma _{S}^{\rho }(\alpha (x))=t$ implies $%
\sigma _{S}^{\rho }(\beta (x))=t$,

$\alpha (x)\equiv \beta (x)$\quad \textit{iff}$\quad \alpha (x)<\beta (x)$
and $\beta (x)<\alpha (x)$.

We note that the quotient set $\Phi ^{\ast }(x)/\equiv $ is partially
ordered by the order (still denoted by $<$) canonically induced on it by the
preorder $<$. It easy to prove that the poset $(\Phi ^{\ast }(x)/\equiv ,<)$
is a Boolean lattice.

(ii) Let $\alpha (x)\in \Phi ^{\ast }(x)$. We call \textit{physical sentence
associated with} $\alpha (x)$ the (universally) quantified sentence $%
(\forall x)\alpha (x)$, and denote by $\Psi ^{\ast }$ the set of all
physical sentences associated with wffs of $\mathcal{L}^{\ast }(x)$ (hence $%
\Psi ^{\ast }=\{(\forall x)\alpha (x)\mid \alpha (x)\in \Phi ^{\ast }(x)\}$%
). Then, for every $S\in \mathcal{S}$, we introduce a \textit{classical
assignment function }$\sigma _{S}:\Psi ^{\ast }\longrightarrow \{t,f\}$ by
setting, for every physical sentence $(\forall x)\alpha (x)\in \Psi ^{\ast }$%
,

$\sigma _{S}((\forall x)\alpha (x))=t$\quad \textit{iff\quad }for every $%
\rho \in \mathcal{R}$, $\sigma _{S}^{\rho }(\alpha (x))=t$.

The logical preorder and equivalence defined on $\Phi ^{\ast }(x)$ can be
extended to $\Psi ^{\ast }$ in a standard way, as follows.

Let $(\forall x)\alpha (x)$, $(\forall x)\beta (x)\in \Psi ^{\ast }$. Then,

$(\forall x)\alpha (x)<(\forall x)\beta (x)$\quad \textit{iff}\quad for
every $S\in \mathcal{S}$, $\sigma _{S}((\forall x)\alpha (x))=t$ implies $%
\sigma _{S}((\forall x)\beta (x))=t$,

$(\forall x)\alpha (x)\equiv (\forall x)\beta (x)$\quad \textit{iff}$\quad
(\forall x)\alpha (x)<(\forall x)\beta (x)$ and $(\forall x)\beta
(x)<(\forall x)\alpha (x)$.

The quotient set $\Psi ^{\ast }/\equiv $ is partially ordered by the order
(still denoted by $<$) canonically induced on it by the preorder $<$, but
the poset $(\Psi ^{\ast }/\equiv ,<)$ is not bound to be a lattice.

(iii) We use the definitions in (ii) to introduce a notion of \textit{true
with certainty} on $\Phi ^{\ast }(x)$. For every $\alpha (x)\in \Phi ^{\ast
}(x)$ and $S\in \mathcal{S}$, we put

$\alpha (x)$ is \textit{certainly true} in $S$\quad \textit{iff}$\quad
\sigma _{S}((\forall x)\alpha (x))=t$ (equivalently, the physical sentence $%
(\forall x)\alpha (x))$ associated with $\alpha (x)$ is \textit{true}).

A wff $\alpha (x)\in \Phi ^{\ast }(x)$ can be certainly true in the state $S$
or not. It must be stressed that in the latter case we do not say that $%
\alpha (x)$ is \textit{certainly false} in $S$: this term will be introduced
indeed at a later stage,with a different meaning. We also note explicitly
that the new truth value is attributed or not to a wff of $\Phi ^{\ast }(x)$
independently of the interpretation $\rho $.

The notion of true with certainty allows one to introduce a \textit{physical
preorder} $\prec $ and a \textit{physical equivalence} $\approx $ on $\Phi
^{\ast }(x)$, as follows.

Let $\alpha (x)$, $\beta (x)\in \Phi ^{\ast }(x)$. Then,

$\alpha (x)\prec \beta (x)$\quad \textit{iff}\quad for every $S\in \mathcal{S%
}$, $\alpha (x)$ is certainly true in $S$ implies that $\beta (x)$ is
certainly true in $S$ (equivalently, $(\forall x)\alpha (x))<(\forall
x)\beta (x)$).

$\alpha (x)\approx \beta (x)$\quad \textit{iff}$\quad \alpha (x)\prec \beta
(x)$ and $\beta (x)\prec \alpha (x)$ (equivalently, $(\forall x)\alpha
(x)\equiv (\forall x)\beta (x)$).

It is apparent that the logical preorder $<$ and the logical equivalence $%
\equiv $ on $\Phi ^{\ast }(x)$ imply the physical preorder $\prec $ and the
physical equivalence $\approx $, respectively, while the converse
implications generally do not hold. Moreover, one can introduce the quotient
set $\Phi ^{\ast }(x)/\approx $, partially ordered by the order (still
denoted by $\prec $) canonically induced on it by the preorder $\prec $
defined on $\Phi ^{\ast }(x)$. Then, the posets $(\Phi ^{\ast }(x)/\approx
,\prec )$ and $(\Psi ^{\ast }/\equiv ,<)$ are obviously order-isomorphic.

(iv) We want to introduce a concept of \textit{testability} on $\Phi ^{\ast
}(x)$. To this end, let us consider an elementary wff $E(x)\in \Phi ^{\ast
}(x)$ and observe that it is testable in the sense that its truth value for
a given interpretation $\rho $ and state $S$ can be empirically checked by
using one of the registering devices in the class denoted by $E$ in order to
perform a registration on $\rho _{S}(x)$. Let us consider now a molecular
wff $\alpha (x)$ of $\Phi ^{\ast }(x)$ and agree that it is testable iff a
registering device exists that allows us to check its truth value. Since we
have assumed that every registering device belongs to an effect, we conclude
that $\alpha (x)$ is testable iff it is logically equivalent to an
elementary wff of $\Phi ^{\ast }(x)$. Thus, we introduce the subset $\Phi
_{T}^{\ast }(x)$ of all testable wffs of $\Phi ^{\ast }(x)$, defined as
follows.

$\Phi _{T}^{\ast }(x)=\{\alpha (x)\in \Phi ^{\ast }(x)\mid \exists E_{\alpha
}\in \mathcal{E}^{\ast }:\alpha (x)\equiv E_{\alpha }(x)\}$.

Of course, the binary relations $<$, $\equiv $, $\prec $ and $\approx $
introduced on $\Phi ^{\ast }(x)$ can be restricted to $\Phi _{T}^{\ast }(x)$%
, and we still denote these restrictions by the symbols $<$, $\equiv $, $%
\prec $ and $\approx $, respectively, in the following.

(v) The notion of testability can be extended to the physical sentences
associated with wffs of $\Phi ^{\ast }(x)$ by setting, for every $\alpha
(x)\in \Phi ^{\ast }(x)$,

$(\forall x)\alpha (x)$ is testable\quad \textit{iff}$\quad \alpha (x)$ is
testable (equivalently, $\alpha (x)\in \Phi _{T}^{\ast }(x)$).

We denote the set of all testable physical sentences by $\Psi _{T}^{\ast }$
(hence, $\Psi _{T}^{\ast }=\{(\forall x)\alpha (x)\mid \alpha (x)\in \Phi
_{T}^{\ast }(x)\}$), and still denote the restrictions to $\Psi _{T}^{\ast }$
of the binary relations $<$ and $\equiv $ defined on $\Psi ^{\ast }$ by $<$
and $\equiv $, respectively. It is then easy to show that the posets $(\Phi
_{T}^{\ast }(x)/\approx ,\prec )$ and $(\Psi _{T}^{\ast }/\equiv ,<)$ are
order-isomorphic.

\section{Physical propositions}

Let $\alpha (x)\in \Phi ^{\ast }(x)$. We put

$p_{\alpha }^{f}=\{S\in \mathcal{S\mid }\alpha (x)$ is certainly true in $%
S\} $,

\noindent
and say that $p_{\alpha }^{f}$ is the \textit{physical proposition
associated with }$\alpha (x)$ (or, briefly, the \textit{physical proposition
of }$\alpha (x)$). It is then easy to see that $p_{\alpha }^{f}$ is the
proposition associated with $(\forall x)\alpha (x)$ according to the
standard rules of a Kripkean semantics in which states play the role of
possible worlds. More formally,

$p_{\alpha }^{f}=\{S\in \mathcal{S\mid }\sigma _{S}((\forall x)\alpha
(x))=t\}=\{S\in \mathcal{S\mid }$for every $\rho \in \mathcal{R}$, $\sigma
_{S}^{\rho }(\alpha (x))=t\}$.

We denote by $\mathcal{P}^{\ast f}$ the set of all physical propositions of
wffs of $\Phi ^{\ast }(x)$,

$\mathcal{P}^{\ast f}=\{p_{\alpha }^{f}\mid \alpha (x)\in \Phi ^{\ast }(x)\}$%
.

The definitions of certainly true in $S$, physical order $\prec $ and
physical equivalence $\approx $ can be restated by using the notion of
physical proposition. Indeed, for every $\alpha (x)$, $\beta (x)\in \Phi
^{\ast }(x)$,

$\alpha (x)$ is certainly true in $S$\quad \textit{iff}\quad $S\in p_{\alpha
}^{f}$,

$\alpha (x)\prec \beta (x)$\quad \textit{iff}\quad $p_{\alpha }^{f}\subseteq
p_{\beta }^{f}$,

$\alpha (x)\approx \beta (x)$\quad \textit{iff}$\quad p_{\alpha
}^{f}=p_{\beta }^{f}$.

The above results imply that the posets $(\Phi ^{\ast }(x)/\approx ,\prec )$
(or $(\Psi ^{\ast }/\equiv ,<)$) and $(\mathcal{P}^{\ast f},\subseteq )$ are
order-isomorphic.\footnote{%
This isomorphism suggests that one could introduce the notion of \textit{%
true with certainty} by firstly assigning $(\mathcal{P}^{\ast f},\subseteq )$
with its algebraic structure and then connecting it with $\Phi ^{\ast }(x)$,
thus providing an \textit{algebraic semantics} which allows one to avoid the
definition introduced in Sec. 2, hence the introduction of a classical truth
theory. One would thus follow standard procedures in QL, yet losing the
links between two different notions of truth illustrated in this paper.}
However, the set-theoretical operations on $\mathcal{P}^{\ast f}$ do not
generally correspond to logical operations on $\Phi ^{\ast }(x)$. Indeed,
for every $\alpha (x)$, $\beta (x)$, $\gamma (x)\in \Phi ^{\ast }(x)$, one
gets

$\alpha (x)\equiv \lnot \beta (x)$\quad \textit{implies}\quad $p_{\alpha
}^{f}\subseteq \mathcal{S\setminus }p_{\beta }^{f}$,

$\alpha (x)\equiv \beta (x)\wedge \gamma (x)$\quad \textit{implies}\quad $%
p_{\alpha }^{f}=p_{\beta }^{f}\cap p_{\gamma }^{f}$,

$\alpha (x)\equiv \beta (x)\vee \gamma (x)$\quad \textit{implies}\quad $%
p_{\alpha }^{f}\supseteq p_{\beta }^{f}\cup p_{\gamma }^{f}$

\noindent
(see also Garola and Sozzo, 2006).

Let us consider now the subset $\Phi _{T}^{\ast }(x)$ of all testable wffs
of $\Phi ^{\ast }(x)$ introduced in Sec. 2. We define the subset $\mathcal{P}%
_{T}^{\ast f}\subseteq \mathcal{P}^{\ast f}$of all testable physical
propositions by setting

$\mathcal{P}_{T}^{\ast f}=\{p_{\alpha }^{f}\mid \alpha (x)\in \Phi
_{T}^{\ast }(x)\}$.

Then, one gets that $\mathcal{P}_{T}^{\ast f}$ coincides with the set of all
physical propositions associated with elementary wffs of $\Phi ^{\ast }(x)$.
Moreover, the posets $(\Phi _{T}^{\ast }(x)/\approx ,\prec )$ (or $(\Psi
_{T}^{\ast }/\equiv ,<)$) and $(\mathcal{P}_{T}^{\ast f},\subseteq )$ are
order-isomorphic.

\section{The language of properties $\mathcal{L}(x)$}

Both in CM and in QM the set of all effects contains a subset of \textit{%
decision effects} (see, e.g., Ludwig, 1983, Ch. III) that we briefly call 
\textit{properties} in this paper. Hence the set $\mathcal{E}^{\ast }$ of
all predicates of $\mathcal{L}^{\ast }(x)$ contains a subset $\mathcal{E}$
of predicates denoting properties. Therefore one can consider the
sublanguage $\mathcal{L}(x)$ of $\mathcal{L}^{\ast }(x)$ constructed by
using only predicates in $\mathcal{E}$ and following the procedures
summarized in Sec. 2. Thus, the set of all wffs of $\mathcal{L}(x)$, the set
of all elementary wffs of $\mathcal{L}(x)$, the semantics and the physical
interpretation of $\mathcal{L}(x)$, the logical preorder and equivalence on $%
\mathcal{L}(x)$, etc., are defined as in Sec. 2, simply dropping the suffix $%
^{\ast }$. Hence one obtains that the poset $(\Phi (x)/\equiv ,<)$ is a
Boolean lattice and that the posets $(\Phi (x)/\approx ,\prec )$ and $(\Psi
/\equiv ,<)$ are order-isomorphic. Moreover, the set $\Phi _{T}(x)$ of all
testable wffs of $\mathcal{L}(x)$ is defined as follows,

$\Phi _{T}(x)=\{\alpha (x)\in \Phi (x)\mid \exists E_{\alpha }\in \mathcal{E}%
:\alpha (x)\equiv E_{\alpha }(x)\}$,

\noindent
and the posets $(\Phi _{T}(x)/\approx ,\prec )$ and $(\Psi _{T}/\equiv ,<)$
are order-isomorphic. It must be noted, however, that the notion of
testability introduced in this way on $\Phi (x)$ does not coincide with the
notion of testability following from the general definition in Sec. 2.
Indeed, according to the latter, the set of all testable wffs of $\Phi (x)$
would be given by

$\Phi _{T}^{\prime }(x)=\{\alpha (x)\in \Phi (x)\mid \exists E_{\alpha }\in 
\mathcal{E}^{\ast }:\alpha (x)\equiv E_{\alpha }(x)\}$,

\noindent
which implies $\Phi _{T}(x)\subseteq \Phi _{T}^{\prime }(x)$, so that $\Phi
_{T}(x)$ and $\Phi _{T}^{\prime }(x)$ cannot, in general, be identified.
Therefore we call \textit{p-testability} the more restrictive notion of
testability introduced here. We notice that the broadening of the set of
testable wffs of $\Phi (x)$ following from considering the language of
effects illustrates from our present viewpoint one of the known advantages
of unsharp QM with respect to standard QM. Exploring this topic goes,
however, beyond the scopes of the present paper.

Let us come now to propositions. The set $\mathcal{P}^{f}$ of all physical
propositions associated with wffs of $\Phi (x)$ can be defined as in Sec. 3,
replacing $\Phi ^{\ast }(x)$ by $\Phi (x)$. Again, no change is required,
but dropping the suffix $^{\ast }$. Hence, proceeding as in Sec. 3, one can
show that the posets $(\Phi (x)/\approx ,\prec )$ (or $(\Psi /\equiv ,<)$)
and $(\mathcal{P}^{f},\subseteq )$ are order-isomorphic. One can then
introduce the subset

$\mathcal{P}_{T}^{f}=\{p_{\alpha }^{f}\in \mathcal{P}^{f}\mid \alpha (x)\in
\Phi _{T}(x)\}\subseteq \mathcal{P}^{f}$

\noindent
of all p-testable physical propositions and the subset

$\mathcal{P}_{T}^{f\prime }=\{p_{\alpha }^{f}\in \mathcal{P}^{f}\mid \alpha
(x)\in \Phi _{T}^{\prime }(x)\}\subseteq \mathcal{P}^{f}$

\noindent
of all testable physical propositions (with $\mathcal{P}_{T}^{f}\subseteq 
\mathcal{P}_{T}^{f\prime }$). The distinction between $\mathcal{P}_{T}^{f}$
and $\mathcal{P}_{T}^{f\prime }$ is relevant in principle. However, we are
only concerned with the subset $\mathcal{P}_{T}^{f}$ in the following. One
easily gets, proceeding as in Sec. 3, that $\mathcal{P}_{T}^{f}$ coincides
with the set of all physical propositions associated with elementary wffs of 
$\Phi (x)$, and that the posets $(\Phi _{T}(x)/\approx ,\prec )$ (or $(\Psi
_{T}/\equiv ,<)$) and $(\mathcal{P}_{T}^{f},\subseteq )$ are
order-isomorphic.

\section{Physical propositions in classical mechanics}

One can consider specific physical theories within the general scheme worked
out in Secs. 2-4 by inserting in it suitable assumptions suggested by the
intended interpretation in Sec. 2. In the case of CM, this leads to the
collapse of a number of notions, which explains why some relevant conceptual
differences have been overlooked in classical physics. Let us discuss
briefly this issue.

First of all, all physical objects in a given state $S$ possess the same
properties according to CM. This feature can be formalized by introducing
the following assumption.

\smallskip

CMS. The set $\mathcal{E}$ of all properties is such that, for every $E\in 
\mathcal{E}$ and $S\in \mathcal{S}$, either $ext_{S}E=\mathcal{U}_{S}$ or $%
ext_{S}E=\mathcal{\emptyset }$.

\smallskip

Let us consider the language $\mathcal{L}(x)$ in CM. Because of axiom CMS,
the restriction of the assigment function $\sigma _{S}^{\rho }$ to $\Phi (x)$
does not depend on $\rho $, hence for every state $S$ the wff $\alpha (x)$
is true iff the physical sentence $(\forall x)\alpha (x)$ associated with it
is true. Thus, the notions of \textit{true} and \textit{certainly true}
coincide on $\Phi (x)$. Hence, the logical preorder and equivalence on $\Phi
(x)$ can be identified with the physical preorder and equivalence,
respectively, so that the posets $(\Phi (x)/\approx ,\prec )$ and $(\Psi
/\equiv ,<)$ can be identified with the Boolean lattice $(\Phi (x)/\equiv
,<) $. Furthermore, all these posets are order-isomorphic to $(\mathcal{P}%
^{f},\subseteq )$, which therefore is a Boolean lattice.

Secondly, let us consider p-testability. It is well known that, in
principle, CM assumes that all properties can be simultaneously tested. This
suggests one to introduce a further assumption, as follows.

\smallskip

CMT. The set $\Phi _{T}(x)$ of all p-testable wffs of $\Phi (x)$ coincides
with $\Phi (x)$.

\smallskip

The above assumption implies $\Psi _{T}=\Psi $ and $\mathcal{P}_{T}^{f}=%
\mathcal{P}^{f}$. Hence, $(\mathcal{P}_{T}^{f},\subseteq )$ is a Boolean
lattice, which explains the common statement that ``the logic of a classical
mechanical system is a classical propositional logic'' (R\'{e}dei, 1998, Ch.
5). However, this statement can be misleading, since it ignores a number of
conceptual distinctions that we have pointed out in our general framework.

\section{Physical propositions in quantum mechanics}

Assumption CMS does not hold in (standard, Hilbert space) QM. Indeed, if $E$
denotes a property and $S$ a state of the physical system $\Sigma $, the
probability of getting result 1 (or 0) when performing a registration by
means of a device belonging to $E$ on a sample of $\Sigma $ may be different
both from 0 and from 1 in QM, which implies (via intended physical
interpretation) that $\emptyset \neq ext_{S}E\neq \mathcal{U}_{S}$. Hence,
one cannot conclude, as in CM, that $(\mathcal{P}^{f},\subseteq )$ is a
Boolean lattice. Moreover, there are properties in QM that cannot be
simultaneously tested. Thus, neither assumption CMT holds, and one cannot
assert that the sets $\mathcal{P}_{T}^{f}$ and $\mathcal{P}^{f}$ coincide.
In order to discuss the order structure of $(\mathcal{P}_{T}^{f},\subseteq )$
in QM, let us firstly introduce the symbols and notions that will be used in
the following.

$\mathcal{H}$: the \textit{Hilbert space} on the complex field associated
with $\Sigma $.

$(\mathcal{L}(\mathcal{H}),\subseteq)$ (briefly, $\mathcal{L}(\mathcal{H})$): the complete, orthomodular, atomic lattice (which also has the covering
property; see, e.g., Beltrametti and Cassinelli, 1981, Ch. 10) of all 
\textit{closed subspaces} of $\mathcal{H}$.

$^{\bot }$, $\Cap$ and $\Cup$: the \textit{orthocomplementation}, the 
\textit{meet} and the \textit{join}, respectively, defined on $\mathcal{L}(\mathcal{H})$.

$\mathcal{A}$: the set of all \textit{atoms} (one-dimensional subspaces) of $%
\mathcal{L}(\mathcal{H})$.

$\varphi $: the bijective mapping $\mathcal{S}\longrightarrow \mathcal{A}$
of all (pure) states on the atoms of $\mathcal{L}(\mathcal{H})$.

$\chi $: the bijective mapping $\mathcal{E\longrightarrow L}(\mathcal{H})$
of all properties on the closed subspaces of $\mathcal{L}(\mathcal{H})$.

$\prec $ : the order on $\mathcal{E}$ canonically induced, via $\chi $, by
the order defined on $\mathcal{L}(\mathcal{H})$.

$^{\bot }$: the orthocomplementation on $\mathcal{E}$ canonically induced,
via $\chi $, by the orthocomplementation defined on $\mathcal{L}(\mathcal{H}%
) $.

The mapping $\chi $ is an order isomorphism of $(\mathcal{E},\prec )$ onto $(%
\mathcal{L}(\mathcal{H}),\subseteq )$ that preserves the
orthocomplementation, hence $(\mathcal{E},\prec )$ also is a complete,
orthomodular, atomic lattice. We call it the \textit{lattice of properties}
of $\Sigma $, and identify it with a (standard, sharp) QL. We then introduce
a further mapping

$\theta $: $E\in \mathcal{E}\longrightarrow \mathcal{S}_{E}=\{S\in \mathcal{S%
}\mid \varphi (S)\subseteq \chi (E)\}\in \mathcal{P}(\mathcal{S})$

\noindent
(where $\mathcal{P}(\mathcal{S})$ denotes the power set of $\mathcal{S}$)
that associates every property $E\in \mathcal{E}$ with the set of states
that are represented by atoms included in the subspace $\chi (E)$. Let $%
\mathcal{L}(\mathcal{S})$ be the range of $\theta $. It is easy to see that
also $(\mathcal{L}(\mathcal{S}),\subseteq )$ is a lattice, isomorphic to $(%
\mathcal{E},\prec )$ and $(\mathcal{L}(\mathcal{H}),\subseteq )$. We still
denote by $^{\bot }$ the orthocomplementation on $(\mathcal{L}(\mathcal{S}%
),\subseteq )$ canonically induced, via $\theta $, by the
orthocomplementation $^{\bot }$ defined on $(\mathcal{E},\prec )$, and call $%
(\mathcal{L}(\mathcal{S}),\subseteq )$ \textit{the lattice of all }$^{\perp
} $\textit{-closed subsets of }$\mathcal{S}$ (for, if $\mathcal{S}_{E}\in 
\mathcal{L}(\mathcal{S})$, $(\mathcal{S}_{E}^{\perp })^{\bot }=S_{E}$).

The interpretations of $(\mathcal{E},\prec )$ and $(\mathcal{L}(\mathcal{H}%
),\subseteq )$ then suggest identifying $\mathcal{L}(\mathcal{S})$ with the
subset of all p-testable propositions. This can be formalized by introducing
the following assumption.

\smallskip

QMT. Let $\alpha (x)\in \Phi _{T}(x)$, and let $E_{\alpha }\in \mathcal{E}$
be such that $\alpha (x)\equiv E_{\alpha }(x)$. Then, the physical
proposition $p_{\alpha }^{f}$ of $\alpha (x)$ coincides with $\mathcal{S}%
_{E_{\alpha }}$ in QM.

\smallskip

Assumption QMT has some relevant immediate consequences. In particular, it
implies that the equivalence relations $\equiv $ and $\approx $ coincide on $%
\Phi _{T}(x)$.\footnote{%
The coincidence of $\equiv $ and $\approx $ suggests that also the logical
preorder $<$ and the physical preorder $\prec $ may coincide on $\Phi
_{T}(x) $ in QM. Indeed, this coincidence has been introduced as an
assumption within the general formulation of the SR interpretation of QM
(see Garola and Solombrino, 1996a). However, we do not need this assumption
in the present paper.} Indeed, note firstly that the bijectivity of the
mapping $\chi $ entails that two properties $E$, $F\in \mathcal{E}$ coincide
iff they are represented by the same subspace of $\mathcal{L}(\mathcal{H})$,
hence iff $\mathcal{S}_{E}=\mathcal{S}_{F}$. Secondly, consider the wffs $%
\alpha (x)$, $\beta (x)\in \Phi _{T}(x)$ and let $E_{\alpha }$, $E_{\beta
}\in \mathcal{E}$ be such that $\alpha (x)\equiv E_{\alpha }(x)$ and $\beta
(x)\equiv E_{\beta }(x)$. Then, the following sequence of coimplications
holds because of assumption QMT,

$\alpha (x)\approx \beta (x)$\quad \textit{iff}\quad $E_{\alpha }(x)\approx
E_{\beta }(x)$\quad \textit{iff}\quad $p_{\alpha }^{f}=p_{\beta }^{f}$\quad 
\textit{iff}\quad $\mathcal{S}_{E_{\alpha }}=\mathcal{S}_{E_{\beta }}$\quad 
\textit{iff}\quad $E_{\alpha }=E_{\beta }$\quad \textit{iff}\quad $E_{\alpha
}(x)\equiv E_{\beta }(x)$\quad \textit{iff}\quad $\alpha (x)\equiv \beta (x)$%
,

\noindent
which proves our statement.

More important for our aims in this paper, assumption QMT implies that the
poset $(\mathcal{P}_{T}^{f},\subseteq )$ of all p-testable physical
propositions associated with wffs of $\Phi _{T}(x)$ (equivalently, with
elementary wffs of $\Phi (x)$) can be identified in QM with the lattice $(%
\mathcal{L}(\mathcal{S}),\subseteq )$ of all $^{\perp }$-closed subsets of $%
\mathcal{S}$. Hence the posets $(\Phi _{T}(x)/\approx ,\prec )$ and $(%
\mathcal{P}_{T}^{f},\subseteq )$, on one side, and the lattices $(\mathcal{L}%
(\mathcal{S}),\subseteq )$, $(\mathcal{L}(\mathcal{H}),\subseteq )$ and $(%
\mathcal{E},\prec )$, on the other side, are order-isomorphic, and the
isomorphisms preserve the orthocomplementation (on $(\mathcal{L}(\mathcal{S}%
),\subseteq )$, $(\mathcal{L}(\mathcal{H}),\subseteq )$ and $(\mathcal{E}%
,\prec )$) or canonically induce it (on $(\Phi _{T}(x)/\approx ,\prec )$ and 
$(\mathcal{P}_{T}^{f},\subseteq )$). We therefore denote
orthocomplementation, meet and join in all these lattices by the same
symbols (that is, $^{\perp }$, $\Cap $ and $\Cup $, respectively). Then, one
can easily show that, for every $\alpha (x)$, $\beta (x)\in \Phi _{T}(x)$,

$\mathcal{S\setminus }p_{\alpha }^{f}\supseteq (p_{\alpha }^{f})^{\perp }\in 
\mathcal{P}_{T}^{f}$,

$p_{\alpha }^{f}\cap p_{\beta }^{f}=p_{\alpha }^{f}\Cap p_{\beta }^{f}\in 
\mathcal{P}_{T}^{f}$,

$p_{\alpha }^{f}\cup p_{\beta }^{f}\subseteq p_{\alpha }^{f}\Cup p_{\beta
}^{f}\in \mathcal{P}_{T}^{f}$.

We can now state our main result in this section. Indeed, \textit{the
isomorphisms above allow one to recover (standard, sharp) QL as a quotient
algebra of wffs of }$\mathcal{L}(x)$\textit{, identifying it with }$(\Phi
_{T}(x)/\approx ,\prec )$. We stress that this identification has required
four nontrivial steps: $^{(i)}$selecting p-testable wffs inside $\Phi (x)$; $%
^{(ii)}$grouping p-testable wffs into classes of physical rather than
logical equivalence; $^{(iii)}$adopting assumption QMT; $^{(iv)}$identifying 
$(\mathcal{L}(\mathcal{S}),\subseteq )$ and $(\mathcal{E},\prec )$.

The above result shows how the non-Boolean lattice of QL can be obtained
without giving up classical semantics, which was our minimal aim in this
paper. However, we have already seen in the Introduction that it has a
deeper meaning if one accepts the SR interpretation of QM. Yet, it must be
noted that no direct correspondence can be established between the logical
operations on $\Phi (x)$ and the lattice operations of QL. By comparing the
relations established in Sec. 3 and the relations above, one gets indeed
that, for every $\alpha (x)$, $\beta (x)$, $\gamma (x)\in \Phi _{T}(x)$,

$\alpha (x)\equiv \lnot \beta (x)$\quad \textit{implies}\quad $p_{\alpha
}^{f}\subseteq \mathcal{S\setminus }p_{\beta }^{f}\supseteq (p_{\beta
}^{f})^{\perp }$,

$\alpha (x)\equiv \beta (x)\wedge \gamma (x)$\quad \textit{implies}\quad $%
p_{\alpha }^{f}=p_{\beta }^{f}\cap p_{\gamma }^{f}=p_{\beta }^{f}\Cap
p_{\gamma }^{f}$,

$\alpha (x)\equiv \beta (x)\vee \gamma (x)$\quad \textit{implies}\quad $%
p_{\alpha }^{f}\supseteq p_{\beta }^{f}\cup p_{\gamma }^{f}\subseteq
p_{\beta }^{f}\Cup p_{\gamma }^{f}$

\noindent
(see also Garola and Sozzo, 2006).

\section{The quantum language $\mathcal{L}_{TQ}(x)$}

The set $\Phi _{T}(x)$ generally is not closed with respect to $\lnot $, $%
\wedge $ and $\vee $, in the sense that negation, meet and join of testable
wffs may be not testable. However, we can construct a language $\mathcal{L}%
_{TQ}(x)$ whose wffs are testable and whose connectives correspond to
lattice operations of QL, as follows.

(i) Let us take $\Phi _{T}(x)$ (equivalently, the set $\mathcal{E}(x)$ of
all elementary wffs of $\Phi (x)$) as set of elementary wffs, and introduce
three new connectives $\lnot _{Q}$, $\wedge _{Q}$ and $\vee _{Q}$(\textit{%
quantum negation}, \textit{quantum meet} and \textit{quantum join},
respectively) and standard formation rules for \textit{quantum well formed
formulas} (briefly, \textit{qwffs}).

(ii) Let $\Phi _{TQ}(x)$ be the set of all qwffs and let us define an
assigment function $\tau _{S}^{\rho }$ on $\Phi _{TQ}(x)$ based on the
assigment function $\sigma _{S}^{\rho }$ defined on $\Phi (x)$. To this end,
let us consider the wffs $\alpha (x)$, $\beta (x)\in \Phi _{T}(x)$ and let $%
E_{\alpha }$, $E_{\beta }\in \mathcal{E}$ be such that $\alpha (x)\equiv
E_{\alpha }(x)$ and $\beta (x)\equiv E_{\beta }(x)$. Then, for every $\rho
\in \mathcal{R}$ and $S\in \mathcal{S}$, we put

$\tau _{S}^{\rho }(\alpha (x))=\sigma _{S}^{\rho }(\alpha (x)),$

$\tau _{S}^{\rho }(\lnot _{Q}\alpha (x))=t$ (or $f$)$\quad $\textit{iff}$%
\quad \sigma _{S}^{\rho }(E_{\alpha }^{\perp }(x))=t$ (or $f$),

$\tau _{S}^{\rho }(\alpha (x)\wedge _{Q}\beta (x))=t$ (or $f$)$\quad $%
\textit{iff}$\quad \sigma _{S}^{\rho }((E_{\alpha }\Cap E_{\beta })(x))=t$
(or $f$),

$\tau _{S}^{\rho }(\alpha (x)\vee _{Q}\beta (x))=t$ (or $f$)$\quad $\textit{%
iff}$\quad \sigma _{S}^{\rho }((E_{\alpha }\Cup E_{\beta })(x))=t$ (or $f$).

It is apparent that $\lnot _{Q}\alpha (x)$, $\alpha (x)\wedge _{Q}\beta (x)$
and $\alpha (x)\vee _{Q}\beta (x)$ are logically equivalent to wffs of $\Phi
_{T}(x)$. Therefore the above semantic rules can be applied recursively by
considering $\alpha (x)$, $\beta (x)\in \Phi _{TQ}(x)$, which defines $\tau
_{S}^{\rho }$ on $\Phi _{TQ}(x)$. Hence, the notions of logical preorder $<$
and logical equivalence $\equiv $ can be extended to $\Phi _{TQ}(x)$, and
every qwff is logically equivalent to a wff of $\mathcal{E}\left( x\right) $
(hence of $\Phi _{T}(x)$).

(iii) Let us associate a physical sentence $(\forall x)\alpha (x)$ with
every qwff $\alpha (x)\in \Phi _{TQ}(x)$. Hence the notions of \textit{%
certainly true}, \textit{physical preorder }$\prec $ and \textit{physical
equivalence }$\approx $ can be introduced on $\Phi _{TQ}(x)$. Furthermore $%
\equiv $ and $\approx $ coincide on $\Phi _{TQ}(x)$, since they coincide on $%
\Phi _{T}(x)$ (Sec. 6).

(iv) For every $\alpha (x)\in \Phi _{TQ}(x)$, let us define the physical
proposition $p_{\alpha }^{f}=\{S\in \mathcal{S\mid }\alpha (x)$ is certainly
true in $S\}$ of $\alpha (x)$. Then, the set of all physical propositions
associated with qwffs of $\Phi _{TQ}(x)$ coincides with $\mathcal{P}_{T}^{f}$%
. Moreover, the semantic rules established above entail that, for every $%
\alpha (x)$, $\beta (x)$, $\gamma (x)\in \Phi _{TQ}(x)$,

$\alpha (x)\equiv \lnot _{Q}\beta (x)$\quad \textit{iff}\quad $p_{\alpha
}^{f}=(p_{\beta }^{f})^{\perp }$,

$\alpha (x)\equiv \beta (x)\wedge _{Q}\gamma (x)$\quad \textit{iff}\quad $%
p_{\alpha }^{f}=p_{\beta }^{f}\Cap p_{\gamma }^{f}$,\footnote{%
Note that, if $\alpha (x)$, $\beta (x)\in \Phi _{T}(x)$, the second
implication at the end of Sec. 6 shows that the physical proposition of $%
\alpha (x)\wedge \beta (x)$ is identical to the physical proposition of $%
\alpha (x)\wedge _{Q}\beta (x)$, which implies $\alpha (x)\wedge \beta
(x)\approx \alpha (x)\wedge _{Q}\beta (x)$. Yet, one cannot assert in this
case that $\alpha (x)\wedge \beta (x)\equiv \alpha (x)\wedge _{Q}\beta (x)$,
since $\alpha (x)\wedge \beta (x)$ does not necessarily belong to $\Phi
_{T}(x)$. The difference between $\wedge $ and $\wedge _{Q}$was overlooked
in a recent paper (Garola and Sozzo, 2006), and we thank S. Sozzo for
bringing such issue to our attention.}

$\alpha (x)\equiv \beta (x)\vee _{Q}\gamma (x)$\quad \textit{iff}\quad $%
p_{\alpha }^{f}=p_{\beta }^{f}\Cup p_{\gamma }^{f}$.

\noindent
(The proof of these coimplications is straightforward if one preliminarily
notices that, for every $E$, $F$ $\in \mathcal{E}$ the physical propositions
of $E^{\perp }(x)$, $(E\Cap F)(x)$ and $(E\Cup F)(x)$ are $%
(p_{E}^{f})^{\perp }$, $p_{E}^{f}\Cap p_{F}^{f}$ and $p_{E}^{f}\Cup
p_{F}^{f} $, respectively, because of the definitions of $^{\perp }$, $\Cap $
and $\Cup $ on $\mathcal{E}$ and assumption QMT).

We have thus constructed a language $\mathcal{L}_{TQ}(x)$ whose connectives
correspond to lattice operations on QL, as desired. It must be stressed,
however, that the semantic rules for quantum connectives have an empirical
character since they depend on the empirical relations on the set of all
properties, and that these rules coexist with the semantic rules for
classical connectives in our approach.

Finally, we note that, for every $\alpha (x)$, $\beta (x)\in \Phi _{TQ}(x)$,
the following logical equivalence can be proved,

$\alpha (x)\vee _{Q}\beta (x)\equiv \lnot _{Q}((\lnot _{Q}\alpha (x))\wedge
_{Q}(\lnot _{Q}\beta (x)))$,

\noindent
and a \textit{quantum implication} connective $\rightarrow _{Q}$ can be
introduced such that

$\alpha (x)\rightarrow _{Q}\beta (x)\equiv (\lnot _{Q}\alpha (x))\vee
_{Q}(\alpha (x)\wedge _{Q}\beta (x))$.

The formal structure of the above logical equivalences is well known in QL.
The novelty here is that $\alpha (x)$ and $\beta (x)$ are sentences
referring to individual samples of physical objects, while the wffs of
standard QL represent propositions and do not bear this interpretation.

\section{Quantum truth}

The notion of \textit{true with certainty} is defined in Sec. 2 for all wffs
of $\mathcal{L}^{\ast }(x)$. Yet, only testable wffs of $\mathcal{L}^{\ast
}(x)$ can be associated with empirical procedures that allow one to check
whether they are certainly true or not.

For the sake of simplicity, let us restrict here to the sublanguage $%
\mathcal{L}(x)$ of $\mathcal{L}^{\ast }(x)$ and to the subset $\Phi
_{T}(x)\subseteq \Phi _{T}^{\ast }(x)$ of p-testable wffs (Sec. 4). Then,
the notion of \textit{certainly true} can be worked out in QM in order to
define a notion of quantum truth (briefly, \textit{Q-truth}) on $\Phi
_{T}(x) $, as follows.

\smallskip

QT. Let $\alpha (x)\in \Phi _{T}(x)$ and $S\in \mathcal{S}$. We put

$\alpha (x)$ is \textit{Q-true} in $S\quad $\textit{iff}$\quad S\in
p_{\alpha }^{f}$,

$\alpha (x)$ is \textit{Q-false} in $S\quad $\textit{iff}$\quad S\in
(p_{\alpha }^{f})^{\perp }$,

$\alpha (x)$ has no Q-truth value in $S\quad $\textit{iff}$\quad S\in 
\mathcal{S\setminus }p_{\alpha }^{f}\cup (p_{\alpha }^{f})^{\perp }$.

\smallskip

Bearing in mind our definitions and results in Secs. 3, 4 and 6, we get

$\alpha (x)$ is Q-true in $S\quad $\textit{iff}$\quad \alpha (x)$ is
certainly true in $S\quad $\textit{iff}$\quad (\forall x)\alpha (x)$ is true
in $S$\quad \textit{iff}\quad $E_{\alpha }(x)$ is certainly true in $S$\quad 
\textit{iff}\quad $(\forall x)E_{\alpha }(x)$ is true in $S$.

The notion of Q-false has not yet an interpretation at this stage. However,
we get from its definition

$\alpha (x)$ is Q-false in $S$\quad \textit{iff}\quad $E_{\alpha }^{\perp
}(x)$ is certainly true in $S\quad $\textit{iff}$\quad (\forall x)E_{\alpha
}^{\perp }(x)$ is true in $S$.

Let us remind now that, for every $E\in \mathcal{E}$, the property denoted
by $E^{\perp }$ is usually interpreted in the physical literature as the
equivalence class of registering devices obtained by reversing the roles of
the outcomes 1 and 0 in all registering devices in $E$ (we stress that we
are considering properties here, not generic effects). This suggests one to
add the following assumption to our scheme.

\smallskip

QMN. Let $E\in \mathcal{E}$. Then, $E^{\perp }(x)\equiv \lnot E(x)$.

\smallskip

Assumption QMN implies

$\alpha (x)$ is \textit{Q-false} in $S\quad $\textit{iff}$\quad (\forall
x)\lnot E_{\alpha }(x)$ is true in $S\quad $\textit{iff}$\quad (\forall
x)\lnot \alpha (x)$ is true in $S\quad $\textit{iff}$\quad \lnot \alpha (x)$
is certainly true in $S$,

\noindent
hence we say that $\alpha (x)$ is \textit{certainly false }in $S$ iff it is
Q-false in $S$.

The above terminology implies that $\alpha (x)$ has no Q-truth value in $S$
iff $\alpha (x)$ is neither certainly true nor certainly false in $S$. We
also say in this case that $\alpha (x)$ is \textit{Q-indeterminate }in $S$.

It is now apparent that the notions of truth and Q-truth coexist in our
approach. This realizes an \textit{integrated perspective}, according to
which the classical and the quantum notions of truth are not incompatible.
Our approach also explains the ``metaphysical disaster'' mentioned in the
Introduction (Randall and Foulis, 1983) as following from attributing truth
values that refer to quantified wffs of a first order predicate calculus to
open wffs of the calculus itself.

Let us conclude our paper with some additional remarks.

Firstly, the notion of Q-truth introduced above applies to a fragment only
(the set $\Phi _{T}(x)\subseteq \Phi (x)$) of the language $\mathcal{L}(x)$.
If one wants to introduce this notion on the set of all wffs of a suitable
quantum language, one can refer to the language $\mathcal{L}_{TQ}(x)$
constructed in Sec. 7. Then, all qwffs are testable, and definition QT can
be applied in order to define Q-truth on $\mathcal{L}_{TQ}(x)$ by simply
substituting $\Phi _{TQ}(x)$ to $\Phi _{T}(x)$ in it. Again, classical truth
and Q-truth can coexist on $\mathcal{L}_{TQ}$ in our approach.

Secondly, definition QT can be physically justified by observing that most
manuals and books on the foundations of QM introduce (usually implicitly) a
verificationist notion of truth that can be summarized in our present terms
as follows.

\smallskip

QVT. Let $\alpha (x)\in \Phi (x)$ and $S\in \mathcal{S}$. Then, $\alpha (x)$
is true (false) in $S$ iff:

(i) $\alpha (x)$ is testable;

(ii) $\alpha (x)$ can be tested and found to be true (false) on a physical
object in the state $S$ without altering $S$.

\smallskip

It can be proved that the notion of truth introduced by definition QVT and
the notion of Q-truth introduced by definition QT coincide. The proof is
rather simple but requires some use of the theoretical apparatus of QM
(Garola and Sozzo, 2004).

Finally, a further justification of definition QT can be given by noting
that the notion of \textit{true with certainty} translates in our context
the notion of \textit{\ certain}, or \textit{true, }introduced in some
partially axiomatized approaches to QM (as Piron's, 1976).

\bigskip

\textbf{REFERENCES}

\smallskip

Bell, J. S. (1964). On the Einstein-Podolski-Rosen paradox. \textit{Physics }%
\textbf{1}, 195-200.

Bell, J. S. (1966). On the problem of hidden variables in quantum mechanics. 
\textit{Reviews of Modern Physics} \textbf{38}, 447-452.

Beltrametti, E. and Cassinelli, G. (1981). \textit{The Logic of Quantum
Mechanics}. Addison-Wesley, Reading, MA.

Birkhoff, G. and von Neumann, J. (1936). The logic of quantum mechanics. 
\textit{Annals of Mathematics}\textbf{\ 37}, 823-843

Busch, P., Lahti, P. J., and Mittelstaedt, P. (1991). \textit{The Quantum
Theory of Measurement}. Springer-Verlag, Berlin.

Busch, P. and Shimony, A. (1996). Insolubility of the quantum measurement
problem for unsharp observables. \textit{Studies in History and Philosophy
of Modern Physics} \textbf{27B}, 397-404.

Dalla Chiara, M., Giuntini, R., and Greechie, R. (2004). \textit{Reasoning
in Quantum Theory}. Kluwer, Dordrecht.

Garola, C. (1991). Classical foundations of quantum logic. \textit{%
International Journal of Theoretical Physics} \textbf{30}, 1-52.

Garola, C. (1999). Against `paradoxes': A new quantum philosophy for quantum
mechanics. In \textit{Quantum Physics and the Nature of Reality}, D. Aerts
and J. Pykacz, eds., Kluwer, Dordrecht, 103-140.

Garola, C. (2000). Objectivity versus nonobjectivity in quantum mechanics. 
\textit{Foundations of Physics} \textbf{30}, 1539-1565.

Garola, C. (2002). A simple model for an objective interpretation of quantum
mechanics. \textit{Foundations of Physics} \textbf{32}, 1597-1615.

Garola, C. (2005). MGP versus Kochen-Specker condition in hidden variables
theories. \textit{International Journal of Theoretical Physics} \textbf{44},
807-814.

Garola, C. and Pykacz, J. (2004). Locality and measurements within the SR
model for an objective interpretation of quantum mechanics. \textit{%
Foundations of Physics }\textbf{34}, 449-475.

Garola, C. and Solombrino, L. (1996a). The theoretical apparatus of semantic
realism: A new language for classical and quantum physics. \textit{%
Foundations of Physics }\textbf{26}, 1121-1164.

Garola, C. and Solombrino, L. (1996b). Semantic realism versus EPR-like
paradoxes: The Furry, Bohm-Aharonov and Bell paradoxes. \textit{Foundations
of Physics }\textbf{26}, 1329-1356.

Garola, C. and Sozzo, S. (2004). A semantic approach to the completeness
problem in quantum mechanics. \textit{Foundations of Physics }\textbf{34},
1249-1266.

Garola, C. and Sozzo, S. (2006). On the notion of proposition in classical
and quantum mechanics. In \textit{The Foundations of Quantum Mechanics.
Historical Analysis and Open Questions - Cesena 2004,} C. Garola, A. Rossi
and S. Sozzo, eds., World Scientific, Singapore.

Kochen, S. and Specker, E. P. (1967). The problem of hidden variables in
quantum mechanics. \textit{Journal of Mathematical Mechanics} \textbf{17},
59-87.

Jammer, M. (1974). \textit{The Philosophy of Quantum Mechanics}. Wiley, New
York.

Jauch, J. M. (1968). \textit{Foundations of Quantum Mechanics}.
Addison-Wesley, Reading, MA.

Ludwig, G. (1983). \textit{Foundations of Quantum Mechanics I}. Springer,
New York.

Mermin, N. D. (1993). Hidden variables and the two theorems of John Bell. 
\textit{Reviews of Modern Physics }\textbf{65}, 803-815.

Piron, C. (1976). \textit{Foundations of Quantum Physics. }Benjamin,
Reading, MA.

Randall, C. H. and Foulis, D. J. (1983). Properties and operational
propositions in quantum mechanics. \textit{Foundations of Physics} \textbf{13%
}, 843-857.

R\'{e}dei, M. (1998). \textit{Quantum Logic in Algebraic Approach}. Kluwer,
Dordrecht.

\end{document}